\documentclass[preprint,aps,showpacs,preprintnumbers,amsmath,amssymb]{revtex4}

\usepackage{graphicx}
\usepackage{dcolumn}
\usepackage{bm}
\usepackage{eurosym}
\usepackage{color}

\newcommand{\R}{{\mathbb R}}

\def\tilde{\widetilde}
\def\bfo{\begin{eqnarray*} }
\def\efo{\end{eqnarray*} }
\def\ba{\begin{eqnarray*} }
\def\ea{\end{eqnarray*} }
\def\beq{\begin{eqnarray}}
\def\eeq{\end{eqnarray}}

\def\p{\partial}

\def\p{\partial}

\def\R{\mathbb R}

\def\x{{\bf x}}
\def\y{{\bf y}}

\def\ll{\lambda_l}

\def\om{\omega}
\def\omln{\omega_{l,n}}
\def\tomln{\tilde{\om}_{l,n}}
\def\pln{\psi_{l,n}}
\def\tpln{\tilde{\psi}_{l,n}}
\def\nb{\mathbf n}
\def\omlnb{\omega_{l,\nb}}
\def\tomlnb{\tilde{\om}_{l,\nb}}

\def\tplnb{\tilde{\psi}_{l,\nb}}

\def\etal{\emph{et al.}\,}

\def\etal{{\it et al.}}

\begin{document}

\title [Schr\"odinger Optics]{Superdimensional Metamaterial Resonators}

\bigskip

\author{Allan Greenleaf\,${}^*$, Henrik Kettunen, Yaroslav Kurylev,  Matti Lassas and Gunther Uhlmann}

\affiliation{Department of Mathematics,
 University of Rochester, Rochester, NY 14627}
 
\affiliation{Department of Mathematics, University of  Helsinki, FIN-00014, Finland,}

\affiliation{Department of Mathematical Sciences, University College London, Gower Str, London,
WC1E 6BT, UK}

\affiliation{Department of Mathematics, University of  Helsinki, FIN-00014, Finland,}

\affiliation{$\hbox{Department of Mathematics, University of Washington, Seattle, WA 98195}$\\
${}^*$Authors are listed in alphabetical order}

\pacs{84.40.Ba, 78.67.Pt, 42.79.Ry}

\begin{abstract} We propose  a fundamentally new method for  the design of metamaterial arrays, 
valid for any waves modeled by the Helmholtz equation,
including scalar optics and acoustics.
The design and analysis of these devices is  based on eigenvalue and eigenfunction asymptotics of solutions to  Schr\"odinger  wave equations  {with harmonic and degenerate potentials.} 
These resonators behave superdimensionally,
with  a higher local density of eigenvalues and  
greater concentration of waves than expected from the physical dimension, 
e.g.,    planar resonators  function as  3- or higher-dimensional media,
and bulk material as effectively of dimension  4 or higher. 
Applications include antennas with a high density of resonant frequencies and giant focussing,
and are potentially  broadband.

\end{abstract}

\maketitle

{\it 1. Introduction.} The advent of transformation optics  has resulted in numerous theoretical designs allowing extreme manipulation of waves, including cloaks \cite{GLU,Le,PSS}, field rotators \cite{ChenChanRot}, electromagnetic  wormholes \cite{GKLU2} and illusion optics \cite{Illusion}, among many others.  The ongoing development of metamaterials has allowed some of these plans to be implemented in  at least a reasonable approximation to the theoretically perfect ideal \cite{Sch,another}. 
Since a partial differential equation
may describe a variety of physical waves, 
one theoretical transformation optics design   may in principle be implemented for a number of distinct wave phenomena. 
Thus, a Helmholtz equation design can be applied to scalar optics \cite{Le}, 
{electromagnetism in cylindrical geometry},
acoustics \cite{Acoustic}, small amplitude water waves \cite{Water},  and, via homogenization and a  gauge transformation, 
even matter waves in quantum mechanics \cite{Quantum}. 
 The possibility of realizing such a devices then depends on the ability, for the wave type and wavelength of interest, to fabricate  suitable metamaterials   and  assemble them into the array required by the design.

In this Letter, we propose a new method for designing  material parameters having radical effects on waves
{modeled} by  Helmholtz-type equations.
We refer to this approach
as \emph{Schr\"odinger optics}, since
it is based on the { mathematical} behavior of quantum mechanical waves, as governed by the 
Schr\"odinger  equation with a trapping potential. 
(We stress that Schr\"odinger optics is not inherently quantum, and  applies to any wave phenomena modeled by  a Helmholtz equation; the 
designs most easily realizable using currently available metamaterials are for acoustics and  2D polarized EM.)
Different choices of potential in the Schr\"odinger  equation result in different effects on wave propagation, allowing many degrees of freedom. 
We introduce Schr\"odinger optics by examining one part of the parameter space,
describing  designs, both planar
 and bulk, based on the  QM harmonic oscillator in  1D  and its degenerate variants. 
 { These Schr\"odinger optics media are striking for the \emph{superdimensionality} that  they exhibit,} in which power laws for various physical properties mimic those  of a  larger dimension than the physical dimension. 
  We   focus on two such properties:  (i) very high density  of resonant frequencies over finite but large frequency bands, a density much larger than that dictated by Weyl's Law \cite{Tay} for conventional media; and (ii) giant concentration and amplification of  waves, resulting from the focussing  behavior of the rays in the high frequency limit, controlled by the associated sub-Riemannian geometry \cite{subR1,subR3}. 
  
   \begin{figure}[htbp]
\begin{center}
\vspace{-.2cm}
\includegraphics[width=.8\linewidth]{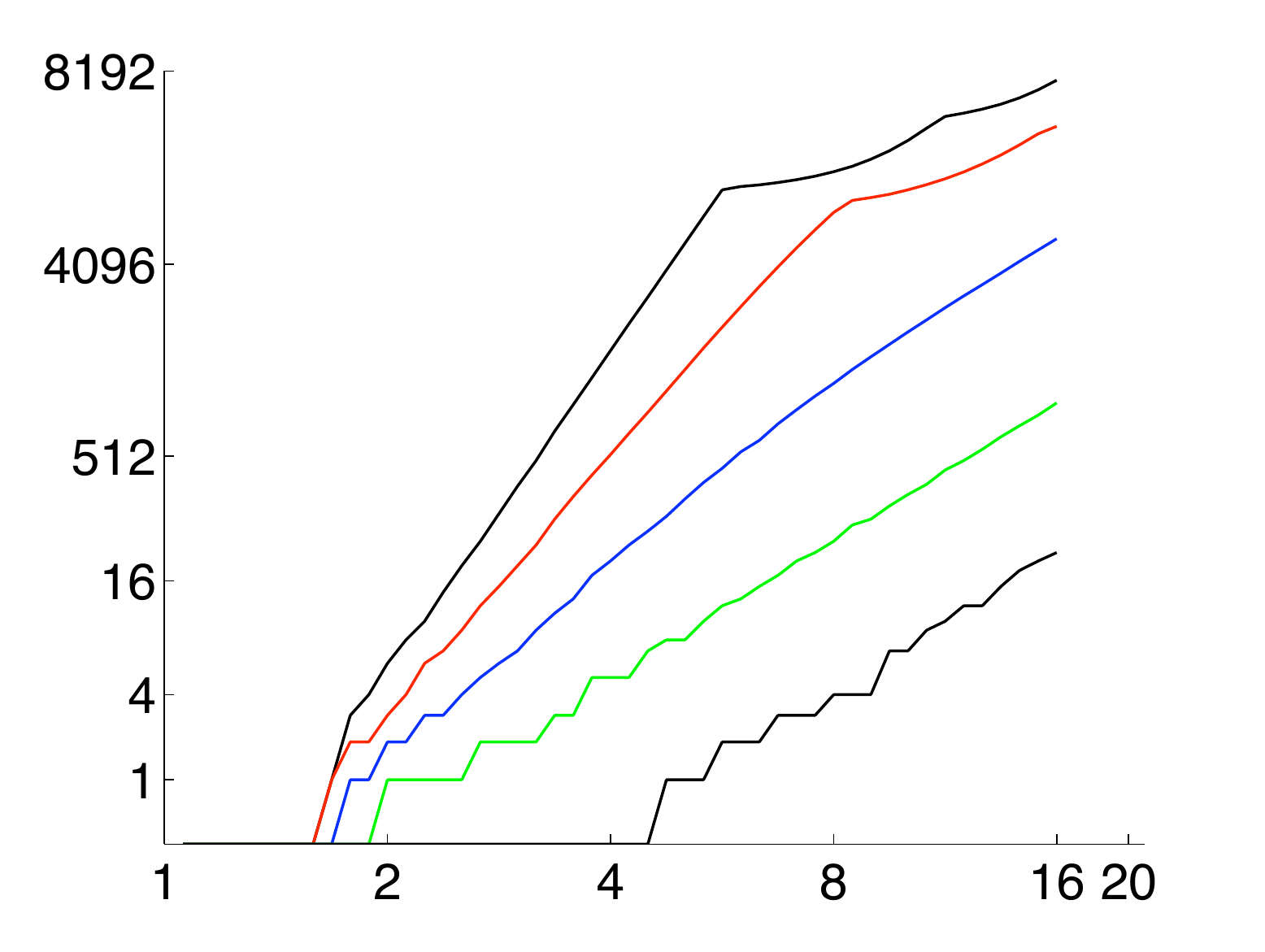}
\end{center}
\vspace{-.8cm}

\caption{{\bf Superdimensional eigenfrequency density.} Log-log graph shows  growth of eigenfrequency count functions $N(\om)$  for five media. 
The eigenfrequency count functions $N(\om)$ 
are shown for approximate SO  resonators
(\ref{eqn approx Grushin}) in the rectangle $R$, with Dirichlet boundary condition,  corresponding to parameters
$a=0.1$ and $r=4$ (upper black curve), $r=3$ (red curve), $r=2$  (blue curve), and $r=1$  (green curve). The  eigenfrequency count function for
the homogeneous material  corresponding to $r=0$ is shown in
the lower black curve. The  eigenfrequency count functions  
display the are superdimensional behavior for $0<\om<\Omega_r=\sqrt{c_r} a^{-1}$, with high density  of frequencies near $\Omega_r$, and the usual 2-dimensional   behavior in the high frequency regime $\om>\Omega_r$.}
\vspace{-.4cm}
\end{figure}

\vspace{1cm}
  
  {\it 2. Methods.}
Ideal Schr\"odinger optics (SO)  designs require spatially {varying} and anisotropic material  parameters 
with infinitely slow wave propagation at some points and in some directions. Recall that the metamaterials required for many transformation optics  designs involve parameters at or close to zero and  are (for EM)  inherently dispersive, or have negative index and  are  lossy. In contrast,  SO designs, since  they do not require superluminal phase velocity,  are potentially broadband. We   analyze both ideal and approximate SO designs, the latter with the potential to be physically realizable (at the price of a quantifiable degradation in performance),  but with the characteristic features of ideal SO still present.

{\it 2.1. 2D ideal Schr\"odinger optics metamaterial} - We start by considering  model SO designs based on  quantum  oscillators in 1D. Fix an integer $r\ge 1$ and consider  a Helmholtz equation in 2D  at frequency $\om>0$  for a wave $u(\x)$ on a  rectangle $R=\{\x=(x,y): |x|\le 1,\, 0\le y\le 1\}$,
\begin{equation}\label{eqn r-Grushin}
\left(\left(\p^2_x + x^{2r}\p^2_y\right)  +\om^2 \right)u(\x)=0.
\end{equation}
(For $\omega=0$,  this 
equation  was first studied by Grushin \cite{Grushin}.) 
We impose the Dirichlet boundary condition (BC) on the boundary of $R$,
{$u=0$ for $ |x|=1\hbox{ or } y= 0,1$,} representing, e.g., a sound-soft surface in acoustics,
but similar results hold for Neumann or mixed BC. 

(i) {\it Superdimensionality of eigenfrequency count} - For waves propagating in homogeneous material in $d$ dimensions,  modeled by $(\nabla^2+\om^2)u=0$,   Weyl's Law states that {in a $d$-dimensional domain} the number of resonant frequencies $\om_j$  
grows as  $N(\om):=\#\{\om_j: \om_j\le\om\}\sim c\cdot\om^d$ \cite{Tay},
{where $d$ is the dimension and $c$ is a positive constant}; this also holds for general \emph{nondegenerate} media, for which the mass-density or analogous tensor is nonsingular, so that the resulting {  PDE} is elliptic. 
There is also a  mathematical  literature \cite{MenSjo} on spectral asymptotics for  \emph{degenerate}-elliptic equations such as (\ref{eqn r-Grushin}), but here  we derive them directly from eigenvalues and eigenfunctions of  quantum harmonic ($r=1$) and anharmonic ($r\ge 2$) oscillators.

For solutions to  (\ref{eqn r-Grushin}) of   the form $u(x,y)=\psi_n(x)\sin(n\pi y),\, n=1,2,\dots$,
on $|x|\le 1$ the $\psi_n$  satisfy 
\begin{equation}\label{eqn Ln}
L_n\psi_n:=\big(\frac{d^2}{dx^2}-\pi^2n^2 x^{2r}\big)\psi_n=-\om^2\psi_n, \,\psi_n(\pm 1)=0.
\end{equation}
For now,  omit the BC and 
 consider the same equation on the entire real line $\R$.
The operator $L:=d^2/dz^2-z^{2r}$ has
eigenvalues $\{-\ll\}_{l=1}^\infty$ and {$L^2$-normalized} eigenfunctions $\{\phi_l\}_{l=1}^\infty$. 
For $r=1$, $\ll=2l+1$ and  the $\phi_l$ are the Hermite functions \cite{Tay}. 
For $r\ge 2$, $\ll$ and $\phi_l$ are less explicit, but it is known  that $\ll\sim c_r l^{\frac{2r}{r+1}}$ and 
$|\phi_l(z)|\le c_{l,\epsilon} exp\left(-\frac{|z|^{r+1}}{r+1}\left(1-\epsilon\right)\right)$ for any $\epsilon>0$ \cite{Lev}.
Letting 
$$\tpln(x)=(\pi n)^{\frac1{2(r+1)}} \phi_l\left(\left(\pi n\right)^{\frac1{r+1}} x\right),\, |x|\le 1,$$ 
$\tpln$ satisfies the  { ordinary differential equation} in 
(\ref{eqn Ln}) with  $\om=\tomln:=(\pi n)^{\frac1{r+1}}\ll^\frac12\sim c_rn^\frac1{r+1}l^{\frac{r}{r+1}}$, 
but with boundary values $\tpln(\pm 1)$ of magnitude \linebreak
$\le c_{r,\epsilon}exp\left(-\left(\frac{\pi n}{r+1}-\epsilon\right)\right)$, for any $\epsilon>0$. 
By standard  perturbation theory \cite{Kato},  { near $\tpln(x), \tomln$ there exist  exact eigenfunctions and eigenfrequencies $\pln(x)$, $\omln$ satisfying (\ref{eqn Ln}) and which are exponentially  (in $n$) close to}  $\tpln(x), \tomln$; for the  purpose of counting resonant frequencies of (\ref{eqn r-Grushin}), we may thus work with the $\tomln$ to estimate $N(\om)$. Ignoring constants, $n^\frac1{r+1} l^\frac{r}{r+1}\le\om$  {if} $n\le\om^{r+1}/l^r$, and  the constraint $n\ge 1$ forces $l\le\om^{\frac{r+1}r}$, so one has 
$$N(\om)\ge \sum_{l=1}^{\om^{\frac{r+1}r}} \frac{\om^{r+1}}{l^r} =\om^{r+1} \sum_{l=1}^{\om^{\frac{r+1}r}} \frac1{l^r},$$
which is  $\sim \om^2\cdot\log\om$ if $r=1$ and $\sim \om^{r+1}$ if $r\ge 2$. 
In comparison with the  classical Weyl power law  (the growth rate of $N(\om)\sim\om^2$ for a 2D  nondegenerate  medium), 
for $r=1$ this ideal Schr\"odinger optics medium exhibits a {\it logarithmically} greater growth, while for $r\ge 2$, the rate is {\it polynomially} greater and in fact is the same as the Weyl Law for $\nabla^2$  in dimension $r+1$,  { see Fig.\ 1. In summary, the eigenfrequency counting function 
has the same growth rate as for an $(r+1)$-dimensional resonator.}

 \begin{figure}[htbp]
\begin{center}
\vspace{-.4cm}
\includegraphics[width=.48\linewidth]{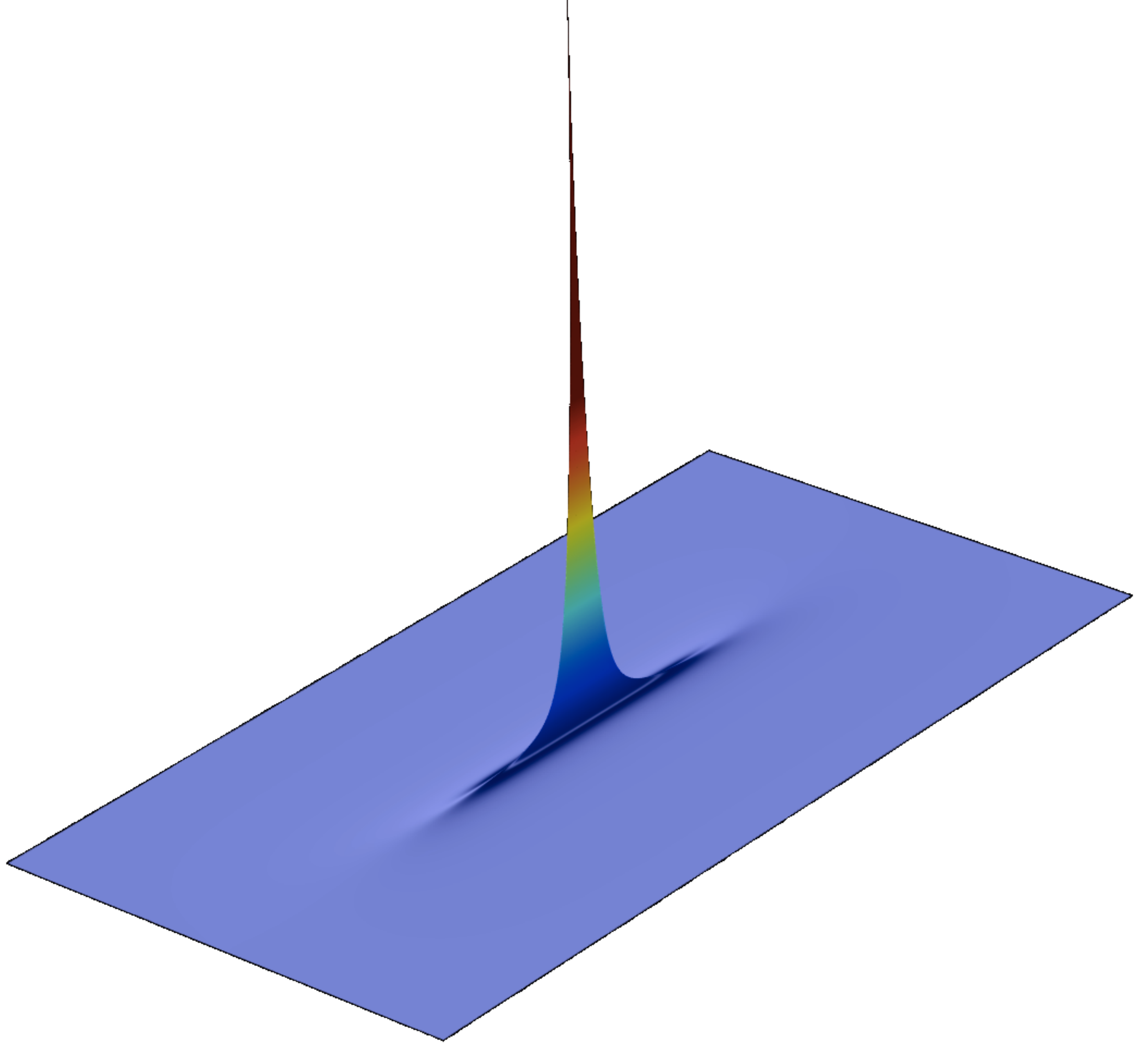}
\includegraphics[width=.48\linewidth]{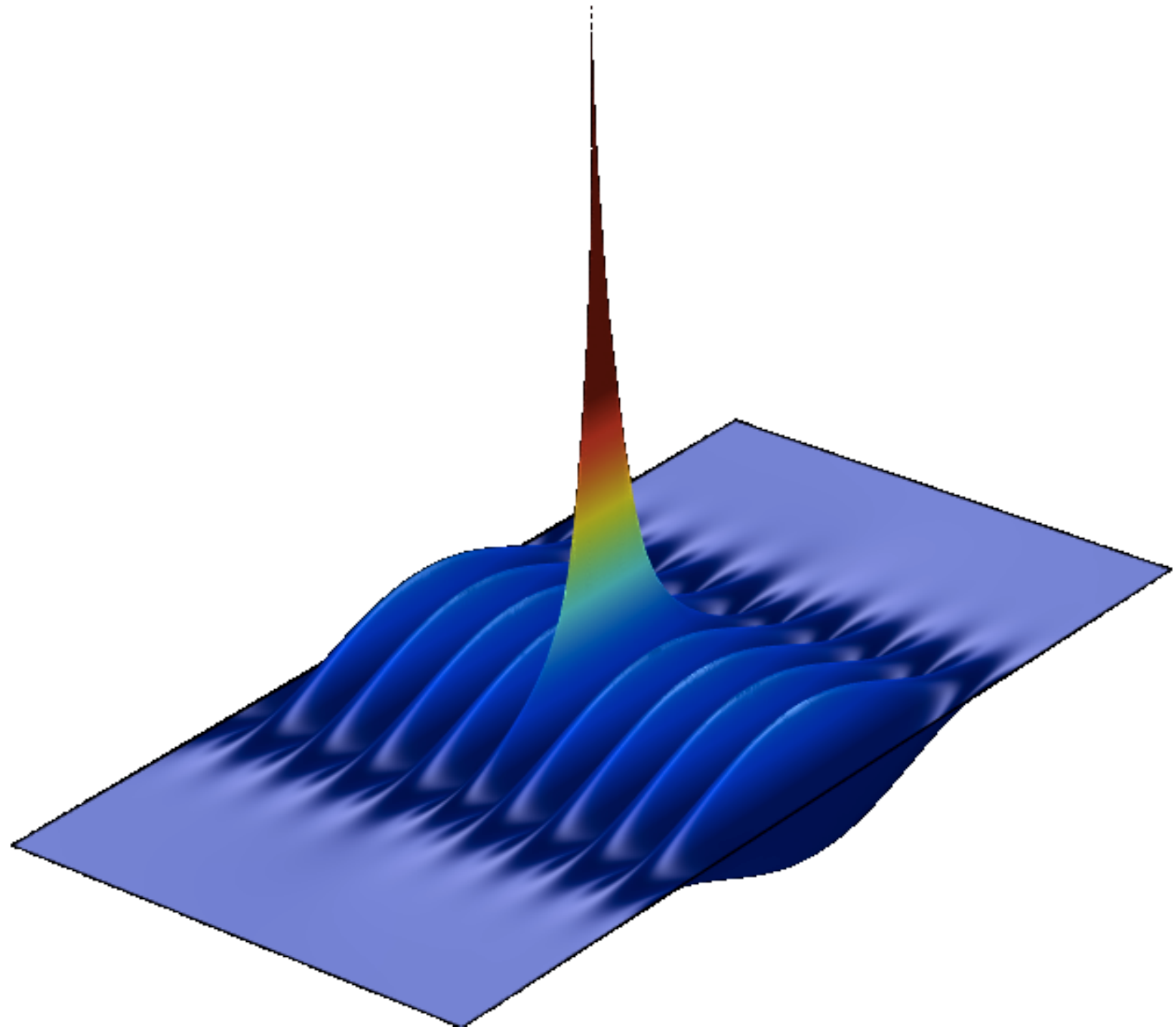}
\vspace{-.6cm}
\end{center}

\caption{\bf Superdimensional wave concentration.
Green's functions for $r=2$ with pole at point $(0,0)$. The   frequency is (left) $\omega=0$ and
 (right) $\omega=4$.
See the \cite{SM} for related figures.
}
\end{figure}

(ii)  {\it Concentration of waves} - SO media also exhibit superdimensionality in the singularity strength  of their Green's functions. For $\om=0$,  the Green's function for (\ref{eqn r-Grushin}) is known analytically  \cite{Beals} and understood from a geometrical point of view for a more general class of equations \cite{RS}.  In 2D, the Newtonian potential   for a homogeneous, isotropic medium ($r=0$), $N(\x,\y)=\frac1{2\pi}\log|\x-\y|$, is both rotation and translation invariant; fixing the pole at $\x=0$,  the singularity is logarithmic and the level curves are circles.   
 In contrast, for $r\ge 1$, the Green's function $G(\x,\y)$ for (\ref{eqn r-Grushin}) reflects both the inhomogeneity of the medium and the degeneracy along $x=0$. Avoiding mathematical details and extracting from the literature only the structure that we need here, write the operator in  (\ref{eqn r-Grushin}) as $X^2+Y^2$, where $X,\,Y$ are the  {first order  differential operators, i.e.,
 vector fields,  $X=\frac \partial{ \partial x},\,  Y=x^r\frac \partial{ \partial y}$.
 The flows of these vector fields define 
  a non-Euclidean metric
 with distance element $ds^2=dx^2+|x|^{-2r}dy^2$  in the rectangle $R$.  
  This  corresponds to {the optical length distance, $D(\x,\y)$, for the wave propagation,
and defines a family of} anisotropic  ``discs",  $B(\x,\delta)$, 
  with center $\x$  and ``radius" $\delta$. 
   The  anisotropic discs are comparable to Euclidean discs away from $x=0$, but those centered along the $y$-axis are comparable to $\delta\times\delta^{r+1}$ rectangles.
 Analysis shows that in a disc  $B({\bf 0},1)$, 
 of radius one centered at the origin, one can pack $C_r\delta^{-(r+1)}$ pairwise disjoint discs
 of radius $\delta$. 
The Hausdorff dimension  \cite{Falc} of  $R$ endowed with the optical length metric is $r+1$, and 
 the SO material behaves as if it were a higher dimensional space.

 We consider the implications of this for the Green's functions.
  {If one denotes by $a(\x,\y)$  the Euclidean  area
  of $B\left(\x,  D\left(\x,\y\right)\right)$, 
the Green's function is known to satisfy \cite{NS}  the estimate
 \begin{equation}\label{eqn estimate}
 \hspace{-1cm}
|G(\x,\y)|\sim c\cdot
\begin{cases} |\log a(\x,\y)| + |\log a(\y,\x)|, \, r=1, \\
\frac{D(\x,\y)^2}{a(\x,\y)} +\frac{D(\y,\x)^2}{ a(\y,\x)},\, r\ge 2.
\end{cases} \hspace{-1cm}
\end{equation}
For $r\ge 2$, and $\x$ on the $y$-axis and $\y=(y_1,y_2)$, one thus has $|G(\x,\y)|\sim (|y_1|+|y_2|^\frac1{r+1})^{-1}$; see  Fig.\ 2. Waves for SO media with point sources on $x=0$ are thus both more singular than the logarithmic blow-up familiar from standard media, and highly concentrated in highly eccentric sets. Similar behavior holds for
$\omega\not =0$ \cite{SM}.

Next consider the high frequency behavior of the Green's functions as $\omega\to \infty$.}
The rays associated to (\ref{eqn r-Grushin}) {have been calculated  \cite{CCGK,ChLi}. Rays through a given point can only point in  directions that are combinations of $X$ and $Y$; thus, passing through $x=0$, motion is only allowed in the $x$ direction.
This causes giant focusing of rays: All rays passing
through a point $P_1$ on that axis $x=0$ focus at $P_1$ so that they have
a common tangent vector at $P_1$ \cite{SM}. Moreover,
for any two points $P_1$  and $P_2$ that are on the axis $x=0$, there are infinitely many rays 
 connecting $P_1$  and $P_2$.  Going from the high frequency limit to a finite frequency $\omega$, one sees that a wave produced by a point source
 at $P_1$ focuses strongly at on many points on  axis $x=0$ (the
 focusing depends on
 the lengths of the connecting rays). This
 yields the strong concentration and large
 oscillation of the waves
 near the axis $x=0$ for large frequencies $\om$ (c.f.\ Fig. 2).}

Devices with anomalous resonant frequency distributions have been described previously, but SO is a fundamentally new approach; e.g., in contrast to fractal antennas \cite{fractal}, SO devices have smooth material parameters and are not self-similar. Designs which strongly  concentrate EM waves have also been obtained previously by transformation optics \cite{LeonLyc}.

{\it 2.2 Approximate Schr\"odinger optics} - 
For  $0<a<1$,
\begin{equation}\label{eqn approx Grushin}
(\p^2_x +(a^{2r}+ x^{2r})\p^2_y) u(\x) +\om^2 u(\x)=0,
\end{equation}
models a medium approximating the ideal SO medium on the rectangle $R$. The maximum degeneracy occurs at $x=0$ and has moderate contrast in the strip $|x|\le a$; outside of this strip the approximate medium is close to the ideal. 
{We  consider  the eigenvalues and eigenfunctions in  $R$
with the  the Dirichlet BC  $u=0$.}
Modifying the analysis above, one sees that  the spectrum for (\ref{eqn Ln}) is shifted by $\pi^2  n^2 a^{2r}$, so that (\ref{eqn approx Grushin}) has eigenfrequencies $\om_{l,n}(a)$ exponentially close to
$$\tomln(a):=(\tomln^2+\pi^2 n^2 a^{2r})^\frac12 \sim (c_r^2n^\frac2{r+1} l^\frac{2r}{r+1} +\pi^2 a^{2r} n^2)^\frac12.$$
For $\tomln(a)\le\om$, we need both $a^{2r}n^2<\om^2$, so that $1\le n\le a^{-r}\om$, and $c_r^2n^\frac2{r+1} l^\frac{2r}{r+1}\le\om^2$, so that $n\ge c_r^{-1}\om^{r+1}/l^r$. In the low frequency regime, $\om<\Omega_r:= (c_r)^{1/r} a^{-1}$, 
the counting function $N(\om,a):= \#\{\om_{l,n}(a): \om_{l,n}(a)\le\om\}$,  
$$N(\om,a)\sim\begin{cases} c_r^{-1}\om^2\log \om, \, r=1,\\ c_r^{-1}\om^{r+1},\, r\ge 2,\end{cases}$$
is superdimensional. For high frequencies, $\om>\nolinebreak\Omega_r$, 
$$N(\om,a)\sim \sum_{l=1}^{c_r^{-\frac1r} a^\frac1r\om} a^{-1}\om + \sum_{l=c_r^{-\frac1r} a^\frac1r\om}^\infty c_r^{-1}\frac{\om^{r+1}}{l^r}\sim d_r a^{\frac1r-1}\om^2,$$
with $d_r=c_r^{-1/r} +\frac{c_r^{(r-1)/r}}{r-1}$, is quadratic. For $r\ge2$ and $\delta>0$, the number of eigenfrequencies in the band $(1-\delta)\Omega_r\le \om\le (1+\delta)\Omega_r$ is $\sim 2c_r^{-1}\delta\Omega_r^{r+1}$, retaining the superdimensionality of the ideal SO medium, while for $\om\to\infty$, the quadratic growth rate of $N(\om,a)$ is as dictated by Weyl's Law in 2D.

 As above, the approximate SO medium corresponds to a frequency shift in Helmholtz for the ideal {SO medium, $\om\to (\om^2+a^{2r})^\frac12$.
When $a$ is small enough,  
  the Green's function is close to the ideal Green's function that has
  a  strong singularity when the source point is  at the surface $x=0$, and for
  high frequencies  exhibits the ideal medium's strong
   concentration of rays.}

{\it 2.3 Implementation} - 
Metamaterial (MM) arrays implementing approximate SO designs at length scale $a<<\nolinebreak1$ can be realized using MM {atoms} composed of  rectangles of low speed material on a substrate  of high speed  material, with increasing vertical fill ratios as $x\to 0$. 
See Fig. 3 and the effective medium theory details in  \cite{SM}.  
The MM atoms used  are nonresonant and thus the 
{standard effective medium approximations
} used to derive the effective parameters should be valid over a broad band.

 \begin{figure}[htbp]
\begin{center}
\vspace{-.2cm}
\hspace{-.5cm}
\includegraphics[width=9cm]{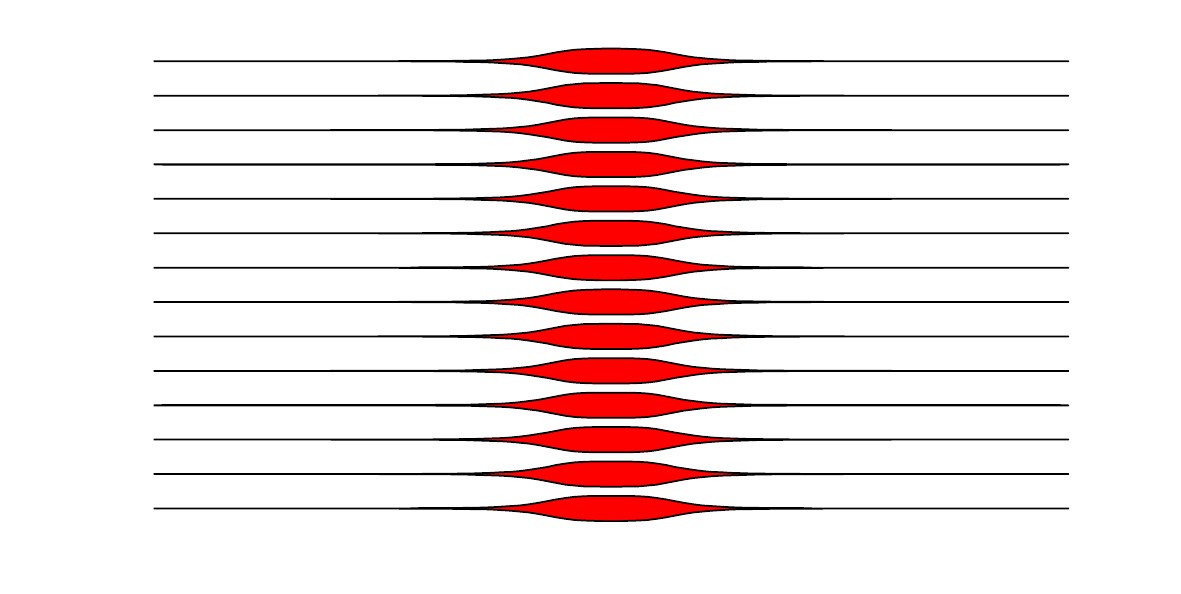}
\vspace{-.4cm}
\end{center}

\caption{{\bf 2D Schr\"odinger optics EM metamaterial.}  
Schematic.  A lattice of sub-wavelength scale $d \times d$ sized unit atoms, each consisting of an $x$-directed strip of varying width $w(x)$ and high permittivity, embedded in background material with relative permittivity. See \cite{SM}.
}
\end{figure}

{\it 2.4. 3D-Schr\"odinger optics} - An ideal bulk SO material in $Q=\{\x=(x,y,z): |x|\le 1, 0\le y\le 1, 0\le z\le 1\}$ may be modeled by replacing (\ref{eqn r-Grushin}) with
$u=0$ on  $\partial Q$ and
\begin{equation}\label{eqn 3DGrushin}
\left(\p^2_x + x^{2r}\left(\p^2_y+\p_z^2\right)  +\om^2 \right)u(\x)=0\hbox{ on } Q.
\end{equation}
For $\nb=(n_1,n_2)\in\mathbb Z_+\times \mathbb Z_+$, eigenfunctions of the form $u(\x)=\psi_{\nb}(x)\sin(\pi n_1 y)\sin(\pi n_2 z)$, $\psi_\nb$ must satisfy (\ref{eqn Ln}) with $n^2$ replaced by $\nb^2:= n_1^2+n_2^2$. As above, one constructs approximate spectral data $\tplnb,\, \tomlnb$, admitting exact solution perturbations $\psi_{l,\nb},\, \omlnb$, exponentially nearby (in $|\nb|$), and to estimate $N(\om)$, it suffices to work with $\tomlnb$. 
As in the 2D case, we need both $|\nb|^\frac1{r+1} l^\frac{r}{r+1}\le\om$, which holds off $|\nb|\le \om^{r+1}/l^r$, and $|\nb|\ge 1$, so that $l\le\om^\frac{r+1}{r}$. Since $\nb$ is 2-dimensional, 
$$N(\om)\ge \sum_{l=1}^{\om^{(r+1)/r}} \frac{\om^{2(r+1)}}{l^{2r}}=c_r\om^{2r+2},$$
so that this 3D ideal SO design exhibits a $(2r+2)$-dimensional eigenfrequency count. For the least degenerate case, $r=1$, this is already 4-dimensional.
As with the 2D case, a more realistic 3D SO design is an approximate one at length scale $a$, with the $x^{2r}$ in (\ref{eqn 3DGrushin}) replaced
by $a^{2r}+x^{2r}$, and this exhibits the superdimensionality in  frequency bands about $\Omega=(c_r)^{1/r} a^{-1}$.

{\it 3. Results} - 
The eigenfrequency counting functions
for the rectangle $R$ with various media have been computed using Matlab, see
Fig.\ 1. The Green's functions have been evaluated using Comsol and the
results are shown in Fig. 2. Documentation of the numerics and
more detailed discussion are in \cite{SM}.

{\it 4. Discussion} - We have shown that metamaterial arrays that exhibit superdimensional behavior can be designed using  the eigenvalue and eigenfunction asymptotics of  
Schr\"odinger operators with trapping potentials. The  examples presented have axial symmetries and are based on harmonic and degenerate harmonic oscillators in 1D. However, the same principles apply to general nonnegative trapping potentials, in one or higher dimensions, with appropriate spectral asymptotics.
{Possible applications include components for antennas that have a
high density of resonance frequencies in a desired frequency band; materials
in which point sources produce fields having extraordinarily strong blow up; and 
optical materials with giant focussing, that either guide light rays together or separate closely propagating rays.}

{\bf Acknowledgments:} AG is supported by  US NSF; YK by  UK EPSRC; HK and ML by Academy of Finland; and GU by  US NSF, a Walker Family Endowed Professorship at UW and a Clay Senior Award.


\begin{thebibliography}{99}

\bibitem{GLU}
A.\ Greenleaf, M.\ Lassas and G.\ Uhlmann, Physiol. Meas. {\bf 24}, 413 (2003); Math. Res. Lett.
{\bf 10}, 685 (2003).

\bibitem{Le} U.\ Leonhardt,
Sci. {\bf 312},
1777 (2006).

\bibitem{PSS}
J.B.\
Pendry, D.\ Schurig, D.R.\ Smith, 
Sci. {\bf 312}, 1780  (2006).

\bibitem{ChenChanRot} H.-Y. Chen, C.T. Chan, 
{\apl} {\bf 90}, 241105 (2007).

\bibitem{GKLU2} A. Greenleaf, Y. Kurylev, M. Lassas, G. Uhlmann,
{\prl}  {\bf 99}, 183901  (2007).

\bibitem{Illusion} Y. Lai, et al., 
\prl  {\bf 102}, 253902 (2009).

\bibitem{Sch}  D. Schurig, \etal, 
Sci. {\bf 314}, 977 (2006).

\bibitem{another} R. Liu, \etal, Sci. {\bf 323}, 366 (2009).

\bibitem{Acoustic} H.-Y. Chen, C.T. Chan, 
\apl {\bf 90}, 241105 (2007); S. Cummer, et al., 
\prl {\bf 100}, 024301 (2008);
A. Greenleaf, Y. Kurylev, M. Lassas, G. Uhlmann, 
http://arXiv.org/abs/0801.3279 (2008).

\bibitem{Water} H.-Y.  Chen, J. Yang, J. Zi, C. T. Chan,
Europhys. Lett.  {\bf 85}, 24004 (2009).

\bibitem{Quantum} S. Zhang, D. Genov, C. Sun,  X.  Zhang,
\prl {\bf 100}, 123002 (2008); A. Greenleaf, Y. Kurylev,  M. Lassas,  G. Uhlmann, 
\prl {\bf 101}, 220404 (2008).


\bibitem{Tay} M. Taylor, Partial Differential Equations, II, Appl. Math. Sci. {\bf 116}, Springer, 2006.

\bibitem{subR1} R. Strichartz, J. Diff. Geom. {\bf 24}, 221  (1986).

\bibitem{subR3} R. Montgomery, A Tour of Subriemannain Geometries, Their Geodesics and Applications, Amer. Math. Soc., Providence, 2002.



\bibitem{Grushin} V. Grushin, Mat. Sb. (N.S.) {\bf 83}, 456 (1970).

\bibitem{MenSjo} A. Menikoff, J. Sj\"ostrand, 
 Math. Annalen {\bf 235}, 55 (1978); C. Fefferman, D. H. Phong, 
Proc. Nat. Acad. Sci. U.S.A. {\bf 77}, 5622  (1980).


\bibitem{Lev}
S.  Levendorskii, 
Russ. Math. 
Surveys {\bf 43}, 148 (1988);
A. Vshivzev, N. Norin, V. Sorokin, Theor. and Math. Phys. {\bf 109}, 139 (1996).

\bibitem{Kato} T. Kato, Perturbation Theory for Linear Operators.  Springer-Verlag. Berlin, 1980.

\bibitem{Beals} R. Beals, Comm. P.D.E. {\bf 24}, 369 (1999); R. Beals, P. Greiner, B. Gaveau, J. Func. Anal. {\bf 165}, 407 (1999).

\bibitem{RS} L. Rothschild, E.M. Stein, Acta Math. {\bf 137}, 247  (1976).

\bibitem{Falc} K. J. Falconer, {The geometry of fractal sets}. Cambridge Tracts in Math. \textbf{85}, Cambridge Univ. Pr., 1986. 

\bibitem{SM} See Supplemental Material  to this paper.


\bibitem{NS} A. Nagel, E.M. Stein, {Lectures on Pseudodifferential Operators},
Princeton Univ. Pr., 1979.


\bibitem{CCGK} O. Calin, D.-C.  Chang, P. Greiner, Y. Kannai, Contemp. Math. {\bf 382}, 89 (2005).

\bibitem{ChLi} D.-C. Chang, Y. Li, 
Jour. Geom. Anal.
{\bf 22}, 800,  (2012).

\bibitem{fractal} D. Werner, S. Ganguly, IEEE Ant. Prop. {\bf 45}, 38 (2003).

\bibitem{LeonLyc} U. Leonhardt, T. Tyc,  New J. Phys. {\bf 10}, 115026 (2008).

\end{thebibliography}
\end{document}